\documentclass{caosp306}
\usepackage{graphicx}

\usepackage{natbib}
\received{November 6, 2018}
\accepted{January 21, 2019}

\def\BibTeX{{\rm B\kern-.05em{\sc i\kern-.025em b}\kern-.08em
             T\kern-.1667em\lower.7ex\hbox{E}\kern-.125emX}}

\begin{document}

\hauthor{P.\,Kab\'{a}th, M.,\,Skarka, S.\,Sabotta and E.\,Guenther}

\title{The role of small telescopes as a ground-based support for exoplanetary space missions}

\author{
        P.\,Kab\'{a}th\inst{1} 
      \and 
        M.\,Skarka\inst{1,2}  
      \and 
        S.\,Sabotta\inst{3}
      \and 
        E.\,Guenther\inst{3}
       }

%%%%%%%%%%%%%%%%%%%%%%%%%%%%%%%%%%%%%%%%%%%%%%%%%%%%%%%%%%%%%%%%%%%%%%%%%%%%%
%           I N S T I T U T E S'  A D D R E S S E S                          
% The affiliation of authors is generated by the \institute command, the
% \and command being again used to separate individual addresses.
% The following commands may be used for the following three institutes:   
%               \lomnica        for      AsU SAV, Tatranska Lomnica          
%               \blava          for      AsU SAV, Bratislava                 
%               \ondrejov       for      AsU CAV, Ondrejov                   
%
% The given postal address must be complete in order to facilitate our
% editorial work. Moreover, you can add your e-mail address, using the
% \email command.
%%%%%%%%%%%%%%%%%%%%%%%%%%%%%%%%%%%%%%%%%%%%%%%%%%%%%%%%%%%%%%%%%%%%%%%%%%%%%
\institute{
           \ondrejov, \email{petr.kabath@asu.cas.cz}
         \and 
           Department of Theoretical Physics and Astrophysics, Masaryk Univerzity, Kotl\'{a}\v{r}sk\'{a} 2, 61137 Brno, Czech Republic 
         \and 
           Th\"{u}ringer Landessternwarte Tautenburg, Sternwarte 5, 07778 Tautenburg, Germany
          }

%%%%%%%%%%%%%%%%%%%%%%%%%%%%%%%%%%%%%%%%%%%%%%%%%%%%%%%%%%%%%%%%%%%%%%%%%%%%%
%                        D A T E / R E C E I V E D                          
% Date inserted here will be the date when your paper was received The
% format is: month (not abbreviated), day, year.
%%%%%%%%%%%%%%%%%%%%%%%%%%%%%%%%%%%%%%%%%%%%%%%%%%%%%%%%%%%%%%%%%%%%%%%%%%%%%
\date{November 6, 2018}
%\date{March 10, 2003}

%%%%%%%%%%%%%%%%%%%%%%%%%%%%%%%%%%%%%%%%%%%%%%%%%%%%%%%%%%%%%%%%%%%%%%%%%%%%%
%                        M A K E T I T L E
% The beginning part (title, author(s), etc.) of your article must be
% closed by the \maketitle command.
%%%%%%%%%%%%%%%%%%%%%%%%%%%%%%%%%%%%%%%%%%%%%%%%%%%%%%%%%%%%%%%%%%%%%%%%%%%%%
\maketitle

%%%%%%%%%%%%%%%%%%%%%%%%%%%%%%%%%%%%%%%%%%%%%%%%%%%%%%%%%%%%%%%%%%%%%%%%%%%%%
%                        A B S T R A C T,  K E Y W O R D S                   
% Here it is shown how to write an abstract.  Keywords should be placed
% within the "abstract" environment using the command \keywords and they
% should be selected from the thesaurus from Astron.  Astrophys.
% Abstracts. They must be separated from each other by -- (two dashes).
%%%%%%%%%%%%%%%%%%%%%%%%%%%%%%%%%%%%%%%%%%%%%%%%%%%%%%%%%%%%%%%%%%%%%%%%%%%%%
\begin{abstract}
Small telescopes equiped with modern instrumentation are gaining on importance, especially, in the era of exoplanetary space missions such as TESS, PLATO and ARIEL. Crucial part of every planet hunting mission is now a ground-based follow-up of detectd planetary candidates. Mid-sized telescopes with apertures of 2 to 4-m with an existing instrumentation become more and more valued due to increasing need for observing time.

In this paper, a brief overview on the follow-up process for exoplanetary space missions will be given. Requirements for the ground-based follow-up instrumentation will be discussed. Some of existing 2-m class telescope facilities and their capability and potential for the follow-up process of exoplanetary candidates will be presented. A special focus will be put on existing 2-m class telescopes in central Europe. 

\end{abstract}

%%%%%%%%%%%%%%%%%%%%%%%%%%%%%%%%%%%%%%%%%%%%%%%%%%%%%%%%%%%%%%%%%%%%%%%%%%%%%
%                       S E C T I O N I N G                                  
% Any section starts with the command \section as shown below, with the
% title in Initial Capitals and lowercase only. Do not number the sections
% - let LaTeX do that for you - and do not end them by a "." (dot).
%
% The (sub)section titles are typeset in boldface; so, if working in the
% mathematics mode in (sub)section titles, you must use \boldmath and 
% enclose it into curly brackets, e.g. "{\bolmath $R^{2}$}".
%%%%%%%%%%%%%%%%%%%%%%%%%%%%%%%%%%%%%%%%%%%%%%%%%%%%%%%%%%%%%%%%%%%%%%%%%%%%%
\section{Introduction}
%%%%%%%%%%%%%%%%%%%%%%%%%%%%%%%%%%%%%%%%%%%%%%%%%%%%%%%%%%%%%%%%%%%%%%%%%%%%%
%                       L A B E L                                            
% The label command is very convenient for you when referring to sections,
% subsections,..., tables, figures as well as to equations (see commands
% \ref and \pageref). In the case of figure and/or table environments the
% \label command should always be put after the \caption command to
% preserve proper numbering. When using the \label command the file must
% be compiled twice to get proper cross-references.
%%%%%%%%%%%%%%%%%%%%%%%%%%%%%%%%%%%%%%%%%%%%%%%%%%%%%%%%%%%%%%%%%%%%%%%%%%%%%
\label{intr}
The first exoplanetary space mission CoRoT was launched in 2006 and was dedicated to asteroseismology and to exoplanets detection \citep{auvergne09}. CoRoT a 28-cm aperture telescope equipped with 4 CCDs, has detected over 30 exoplanets. All CoRoT planets are fully characterized and their masses and radii are known with high accuracy. Furthermore several hundreds of CoRoT candidates are still in the follow-up verification process. Among the first reported exoplanets was also at that time the smallest rocky planet CoRoT-7b \citep{leger07}. However, CoRoT space mission is also a great example to demonstrate the need for a ground-based follow-up for space missions and especially, the need for medium-sized telescopes with up-to-date instrumentation. CoRoT instrument was equipped with prisms which were used with manufactured masks for the selected stars to provide colour information for the target stars. The extra information was useful for analyzing of stellar variability, but the stellar PSF was spread over numerous pixels thus increasing the probability of false positive detections due to blends. Therefore, an extensive follow-up observing was performed with photometric and spectroscopic instruments \citep{deeg09}.

Mid-sized telescopes with apertures up to 4-m with Echelle spectrographs or CCD imagers played an important role in the follow-up process like in case of CoRoT-7 \citep{2011MNRAS.411.1953P}. In the following section, the false positives rejection process will be described. The third section will present an example of mid-sized telescope equipped with Ondrejov Echelle Spectrograph (OES) and Observatory in Ondrejov and how it can help as a ground-based support for space missions and future plans and outlook will be discussed in the Conclusions section.

\section{Source and type of false positives}

Once an exoplanetary candidate is reported from the space mission photometric data, where a transit was detected, the planet verification process begins. The initial task is to determine the stellar parameters such as temperature $T_{\rm eff}$ and stellar radius $R$. For this step, a mid-size telescope with medium resolution spectrographs can be of great help. Furthermore, we can obtain an information about the spectral type of the star when the parameters are combined with e.g. Gaia data. This work can be performed with 2-m class telescopes and their instrumentation.

The next step is to determine the size of the planetary candidate from the combination of spectroscopic and photometric data. But still, the planetary candidate can be a false positive. This could be due to contaminant which is close to the target star and unresolved. Detailed follow-up with high resolution imaging is required. For such observations, typically larger telescopes with AO instruments are used \citep{schmitt16}. In some cases, contaminant stars are detected and the system is marked as a false positive. The transit can also happen on the contaminant system which leads to dillution of the transit depth \citep{ligi18}.

Another step is to determine the radial velocities in two opposite phases to analyze the radial velocity amplitude. From the amplitude determination, it can be decided about the nature of the system because typical radial velocity amplitudes of exoplanetary orgin are below km/s with some exceptions, such as WASP-18b \citep{2009Natur.460.1098H}. If the candidate passes this test, it can be followed-up spectroscopically to determine radial velocities and thus then to obtain mass limit and orbital parameters \citep{bouchy09, loeillet08, gandolfi10, leger07}. For the precise radial velocity measurements a stable instrument on all sizes of telescope apertures can be used. For small planets of Earth-size around solar type stars the required accuracy of radial velocity measurement is in cm/s regime limiting the available instrumentation \citep{2010SPIE.7735E..0FP}. However, Jupiter-sized planets can be followed-up by 2-m class telescopes comfortably, as was done also for the first exoplanet orbiting the solar type star 51 Peg \citep{mayor95}. Later many planets were confirmed with help of 2-m class telescopes \citep{endl02, dollinger07}.

Only after the last described step, the planet is fully characterized and we know its mass and radius. Interesting illustration why the follow-up process is so needed for space missions are the estimates of false positives provided for CoRoT by \cite{almenara09}, for Kepler by \cite{santerne12, fressin13} and for TESS by \cite{santerne13}. In addition, the PLATO space mission team estimated how much time is needed for the spectroscopic ground-based follow-up with different aperture sizes. The amount of time estimated for 2-m class telescopes is about 50 nights per year. In reality, perhaps much more time will be needed. Therefore, a 2-m class telescope with a decent spectrograph will be soon very valuable. Especially, now when the TESS space mission team delivered the first list with candidates and the first planet has been already confirmed \cite{gandolfi18}.

\section{Example of 2-m class facility}

There are several 2-m class telescopes in Europe. The first exoplanet was detected by 1.93-m telescope with the ELODIE instrument at Observatoire de Haute Provence \citep{mayor95, 1996A&AS..119..373B}. Another facility with a 2-m Alfred Jensch telescope at Th\"{u}ringer Landessternwarte is located in Thuringia in Tautenburg, Germany. Another Zeiss built telescope of 2-m class is located at the Rozhen observatory in Bulgaria. %\footnote{http:\/\/nao-rozhen.org\/telescopes\/fr_en.html}. 

\begin{figure}
\centerline{\includegraphics[width=0.95\textwidth]{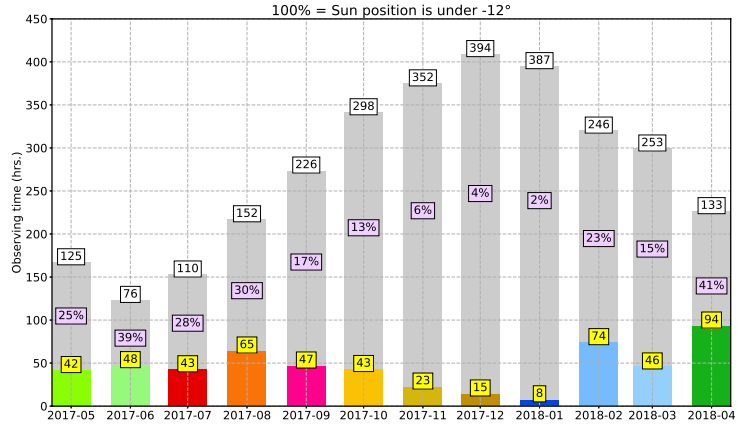}}
\caption{Yearly histogram of available hours of observing for 2017-2018 period (Sun $12^\circ$ below the horizon) versus observed hours. }
\label{fig1}
\end{figure}

\begin{figure}
\centerline{\includegraphics[width=0.95\textwidth]{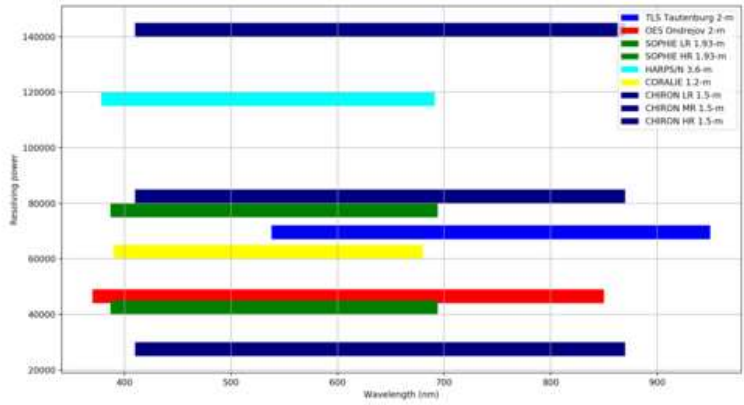}}
\caption{A representative sample of instruments used for exoplanet research on 2-m class telescopes with HARPS/N instruments presented for comparison.}
\label{fig2}
\end{figure}

We present here Perek 2-m telescope located at Ond\v{r}ejov Observatory 30 kilometers southeast from Prague, Czech Republic. Perek telescope was inaugurated as a second Zeiss telescope in 1967 and it is operated by the Czech Academy of Sciences. The location is characterized by typical central European weather and a typical amount of usable telescope time versus observed time is presented in Fig.\ref{fig1}. Usually, the observing time when the telescope was open and science data were obtained is about 20-30 \% of all nights per year when the observing night is defined as with the Sun $12^\circ$ below the horizon. Therefore, the available nights number is between 75-110. The typical seeing value for Ond\v{r}jov is between 2 -- 3 arcsec. The telescope is equipped with two instruments, first being the single order spectrograph and the newer instrument Ondrejov Echelle Spectrograph (OES) with a slit used mainly for spectroscopic ground-based support of exoplanetary missions, such as K2, TESS and later PLATO and ARIEL. Fig.\ref{fig2} presents a comparison of similar instruments on 2-m class telescopes and their wavelength coverage and resolving power with HARPS/N instruments included for illustration.

\begin{table}
\caption{Instrumental characteristics of OES}             % title of Table
\label{tab1}      % is used to refer this table in the text
\centering                          % used for centering table
\begin{tabular}{c c c c}        % centered columns (4 columns)
\hline\hline                 % inserts double horizontal lines
Parameter & value &  \\    % table heading 
\hline                        % inserts single horizontal line
Slit width & 2" on the sky \\
Spectral range & 360--950 nm  \\      % inserting body of the table
Echelle   & 54.5 g/mm  \\
Blaze angle ($\theta$)   &  $69^\circ$ \\
Spectral resolution   & 50,000 (500 nm)  \\
Limiting magnitude & 13 $V$mag \\
\hline\hline                            %inserts single line
\end{tabular}
\end{table}

\begin{figure}
\centerline{\includegraphics[width=0.95\textwidth]{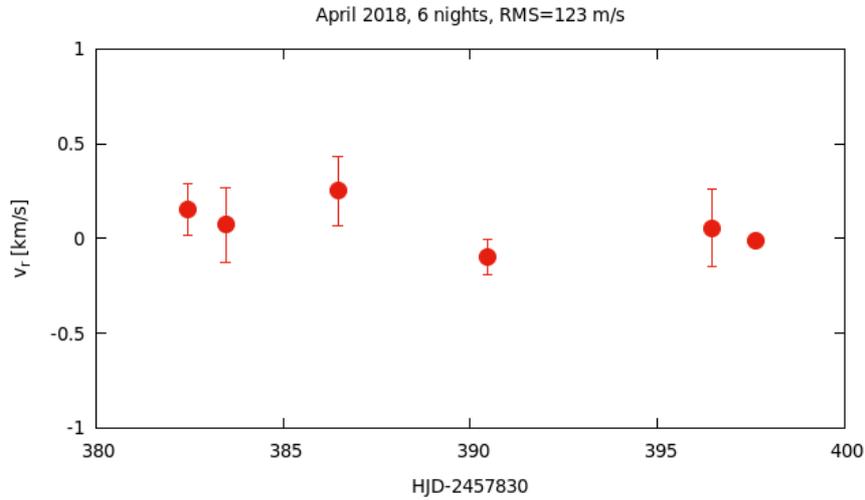}}
\caption{Radial velocity curve for 6 nights of spectroscopic time series for standard HD~109358. Data were taken over a period of 6 nights in April 2018. The RMS corresponds to $123$\,m/s. 
              }
\label{fig3}
\end{figure}

Brief characteristics of OES spectrograph can be found in Tab. \ref{tab1}. OES can monitor a broad range of wavelengths from approximately 360 nm to 950 nm. The accuracy in radial velocities is demonstrated in Fig. \ref{fig3} where measurement of a G-type radial velocity standard star HD109358 (V mag $= 4.25$) can be found. Typical accuracy over one night is better than 100 m/s and even a long term stability of radial velocity measurements spanning over one year is about $300$ m/s, for more detail see \cite{kabath18}, where also first scientific results of false positives rejection are presented. Typically, the performance for star of V mag $=8-9$ of spectral type A is around 1 km/s but this is of course due to the characteristics of the A stars. The limit of OES in terms of magnitudes is $12.5$ mag obtained in 1.5 hrs. exposure with SNR $7$.

These parameters make the OES spectrograph a valuable instrument for the ground-based support of exoplanetary space missions. The reached sensitivity offers a use of OES especially for planetary candidate validation process and characterization of a small sample of hot Jupiters orbiting bright stars. Space mission TESS already delivered the first candidate alert list and the median brightness of the stars in the list is about $10.5$ mag. In the first place, OES will be able to validate and contribute to characterization of a selected sample from TESS exoplanet candidates which will be soon observable from the northern latitudes. The first confirmed TESS exoplanet has $V \rm mag = 5.7$ \citep{gandolfi18}. Therefore, a significant contribution is expected from all 2-m class facilities equipped with spectrographs offering high accuracy in radial velocities. 

\section{Conclusions}

TESS space mission \citep{2015JATIS...1a4003R} was successfully launched in April 2018 and now first exoplanet was confirmed \cite{gandolfi18,huang18} and many more are coming. In the second half of the next decade, the PLATO space mission should be launched in 2026. The PLATO space mission will be searching for transits of extrasolar planets and it will be monitoring the stellar activity of million bright stars up to magnitude 11. According to PLATO red book \citep{red} several thousands of candidates and later confirmed planets are to be expected. Not much later after PLATO another exoplanetary space mission ARIEL should be launched in 2028 \citep{tinetti18}. ARIEL space mission should be searching for signatures of exo-atmospheres, however, a ground-based segment will be needed to support ARIELs operations even well before ARIEL's launch. Observational effort to confirm newly detected exoplanets will be thus growing and demand for observing time on 2-m class facilities will be rapidly increasing.

%%%%%%%%%%%%%%%%%%%%%%%%%%%%%%%%%%%%%%%%%%%%%%%%%%%%%%%%%%%%%%%%%%%%%%%%%%%%%
%                       A C K N O W L E D G E M E N T S                      
% Next lines show you how to write acknowledgements.
% You must leave a blank line before the \acknowledgements command!
%%%%%%%%%%%%%%%%%%%%%%%%%%%%%%%%%%%%%%%%%%%%%%%%%%%%%%%%%%%%%%%%%%%%%%%%%%%%%

\acknowledgements
% Do not leave a blank line here! <---------------------->
PK would like to acknowledge the support from GACR international grant 17-01752J and some travel cost to visit collaborators from ERASMUS+ grant Agreement no. 2017-1-CZ01-KA203-035562. EG and SS acknowledge the DFG grant GU 646/20. MS acknowledges financial support of Postdoc@MUNI project CZ$.02.2.69/0.0/0.0/16\_027/0008360$. We acknowledge the use of IRAF and SIMBAD. We would like to thank to the referee for improving the quality of this paper.

{}

\end{document}